\documentclass[aps,twocolumn,showpacs,amsmath,amssymb,pre,superscriptaddress, floatfix]{revtex4-1}

\usepackage{graphicx}
\usepackage{dcolumn}
\usepackage{bm}
\usepackage{amssymb}
\usepackage{multirow}
\usepackage{braket}

\begin{document}

\title{Symmetries and transport in site-dependently driven quantum lattices}

\author{Thomas Wulf}
    \email{Thomas.Wulf@physnet.uni-hamburg.de}
    \affiliation{Zentrum f\"ur Optische Quantentechnologien, Universit\"at Hamburg, Luruper Chaussee 149, 22761 Hamburg, Germany}
\author{Christoph Petri}

\author{Benno Liebchen}
    \affiliation{Zentrum f\"ur Optische Quantentechnologien, Universit\"at Hamburg, Luruper Chaussee 149, 22761 Hamburg, Germany}

\author{Peter Schmelcher}
    \email{Peter.Schmelcher@physnet.uni-hamburg.de}
    \affiliation{Zentrum f\"ur Optische Quantentechnologien, Universit\"at Hamburg, Luruper Chaussee 149, 22761 Hamburg, Germany}
    \affiliation{The Hamburg Centre for Ultrafast Imaging, Universit\"at Hamburg, Luruper Chaussee 149, 22761 Hamburg, Germany} 

\date{\today}


\begin{abstract}

We explore the quantum dynamics of particles in a spatiotemporally driven lattice. 
A powerful numerical scheme is developed, which provides us with the Floquet modes and thus enables a  
stroboscopic propagation of arbitrary initial states. 
A detailed symmetry analysis represents the cornerstone for an intricate manipulation of the Floquet spectrum.
Specifically, we show how exact crossings can be converted into avoided ones, while the width of these resulting 
avoided crossings can be engineered by adjusting parameters of the local driving.
Asymptotic currents are shown to be controllable over a certain parameter range.

\end{abstract}

\maketitle

\section{Introduction}

Periodically driven systems which allow for the extraction of work out of an unbiased environment are often termed 'ratchets'. Because, originally 
these systems relied on the rectification of thermal noise they were seen as realizations of Brownian motors the corresponding research field being highly active (see \cite{Hanggi:2009} and references therein). 
As one of the many considered experimental playgrounds for ratchets physics, cold atoms loaded into driven optical lattices 
have proven to be particularly insightful since they allow for a precise control over the systems parameters \cite{Renzoni:2006, Renzoni:2010, Renzoni:2012, Salger:2009, Weitz:2013}.
While some of these experiments are carried out at moderate temperatures and allow for a classical treatment \cite{Renzoni:2006, Renzoni:2010, Renzoni:2012}, others could reach ultracold temperatures 
and demonstrated the accessibility of Hamiltonian ratchet setups operating deep in the quantum regime \cite{Salger:2009, Weitz:2013}. 
The experimental advancements concerning these newly realized 'quantum ratchets' were accompanied by a substantial body of theoretical
works (see \cite{Hanggi:2005, Schanz:2001, Schanz:2005, Denisov:2007, Grifoni:2002, Carlo:2005, Sols:2010, Gong:2007} and references therein). 
Thereby, one of the main achievements was the classification of the symmetries which need to be broken in order to allow
for the observation of a ratchet current \cite{Hanggi:2014}. Besides that, the discussed phenomena associated to quantum ratchets were manifold. 
Examples are the existence of resonances in the directed current \cite{Denisov:2007, Flach:2007}, the possibility to tune the dispersion rate of a 
wave packet \cite{Zhan:2011} or the harvesting of Landau-Zener transitions \cite{Gong:2008}. Another active sub-area of ratchet physics are periodically kicked systems,
which have the advantage of being more accessible from a theorists point of view due to the simpler delta-shaped time dependence.
In these, similar effects as for the aforementioned smoothly driven setups could be observed, such as directed transport, resonance behaviour and even 
the acceleration of ratchet currents \cite{Monteiro:2007, Monteiro:2009, Ding:2014, Wimberger:2013, Dana:2008}.\\
In any case, all of the aforementioned works related to ratchet physics, be it classical-, quantum-, smoothly driven- or kicked, are restricted to globally acting driving forces. 
Over the last years however it was shown for classical particles loaded into driven lattices, that a site-dependent driving can add to the already existing diversity of physical phenomena in ratchet 
setups \cite{Petri:2010, Petri:2011, Liebchen:2012, Reimann:2007, Wulf:2012, Wulf:2014, Liebchen:2011}. 
In \cite{Liebchen:2011} for example it was shown how a protocol based 
on a site-dependent driving leads to the patterned deposition of particles out of a uniform initial particle distribution. Furthermore, \cite{Petri:2011, Wulf:2012, Wulf:2014} demonstrated 
the possibility for particles to undergo conversion processes from diffusive- to regular motion, something explicitly forbidden for globally uniform driving forces and which
leads to a plethora of nonequilibrium phenomena for the dynamics. 
Inspired by these
interesting observations, it seems an intriguing perspective to carry over the idea of a spatially non-global driving from the classical setups to the realm of quantum ratchets. This is precisely the purpose of this work.
An additional motivation to this project is the upcoming of cold atom experiments which also deviate from the simple case of a spatially uniform driving force. A noteworthy idea in this context is the introduction 
of sub-wavelength lattices \cite{Weitz:2009} which allow for the construction of more complicated unit cells of the lattice. Full control over each of the lattice barriers is achieved
in experiments with so called 'painted potentials' \cite{Boshier:2009}, even though these are until now restricted to just a few barriers and do not yet consider extended lattices. \\
In the present paper we show that for a quantum particle exposed to a periodically oscillating lattice, the inclusion of a site-dependent driving indeed leads to novel phenomena. 
Such setups are investigated here in the framework of Floquet-Bloch theory and
we demonstrate how, by breaking the translational invariance through the local driving, a set of new symmetry classes for the Floquet-Bloch modes evolves. This is demonstrated to have 
significant impact on the Floquet spectrum where we observe the transformation from exact- to avoided crossings for a deviation from the global driving towards a site-dependent one. Even more, we find that the width 
of the resulting avoided crossing is controllable through variation of a single parameter of the local driving. This is particularly interesting, because the width of avoided crossings in Floquet spectra
of driven lattices is of relevance for a variety of physical phenomena
two noteworthy examples being the already mentioned
Landau-Zener transitions \cite{Flach:2007} or quite generally the diffusion properties of a wave-packet \cite{Graham:1994}. 
Finally, we show the possibility for a directed current in our system, even for the case where all individual barriers 
do not break the relevant symmetries. Consequently, the symmetry breaking becomes a collective phenomenon of the barriers constituting the lattice.\\
This work is structured in the following way: In section \ref{S1} we introduce the setup of the spatiotemporally driven lattice. In section \ref{S2} 
we show how a well-known scheme to calculate the time evolution operator in a periodically driven system can be extended in order to simulate the time evolution 
in a site-dependently driven lattice. In doing so, we also take the possibility of nonzero quasi-momenta into account. In section \ref{S3} we perform a thorough 
symmetry analysis and identify symmetry classes for the Floquet-Bloch modes, which arise
due to the site-dependent driving. In section \ref{S4} we show how the existence of these newly found symmetry classes is translated into properties of 
the Floquet spectrum. Section \ref{S5} contains an investigation of the transport properties of our setup. Finally, we conclude and provide an outlook in section \ref{S6}.

\section{The spatiotemporally driven lattice}
\label{S1}

The system under investigation consists of a single quantum particle in one dimension exposed to a laterally oscillating lattice of Gaussian potential barriers. 
Hence the dynamics obeys the time-dependent Schr\"{o}dinger equation (TDSE)
\begin{equation}
  i \hbar \frac{\partial \Psi(x,t)}{\partial t}=H(x,t) \Psi(x,t)
  \label{Schrodinger}
\end{equation}
where $x$ and $t$ denote position and time and the Hamiltonian is given by
\begin{equation}
  H(x,t)=-\frac{\hbar^2}{2m}  \frac{\partial^2}{\partial x^2} + V_0 \sum_{i=-\infty}^{\infty} e^{-\left(  \frac{x-i\, L- d_i(t)}{\Delta}      \right)^2 }.
\label{Hamiltonian}
\end{equation}
Here, $\Delta$, $L$, $m$ and $V_0$ are the barrier width, the distance between the barriers equilibrium positions, the particles mass and $V_0$ is the height 
of the potential barriers respectively, while $d_i(t)$ is the driving law. 
Without loss of generality we will set $m=\hbar=1$ in the following.
The crucial difference to Hamiltonians usually studied in investigations of
driven lattices in the quantum regime is that the driving law carries 
a barrier index $i$ and thus can be site-dependent. Throughout this work, we employ a cosine driving with equal amplitude and frequency but with possibly different phases:
\begin{equation}
 d_i(t)=A \cos(\omega t + \delta_i).
\label{driving}
\end{equation}
Moreover, we restrict ourselves to sequences of the barrier phases which periodically repeat themselves after some number of barriers $n_p$, i.e. we have $\delta_i=\delta_{i+n_p}$.   
Because of the intimate relation between the driving of a lattice site and its barrier index $i$, and thus its position within the lattice, 
we call a setup with $n_p>1$ a spatiotemporally driven lattice. In contrast, for the case of only a single employed 
driving law, i.e. $n_p=1$, we say the lattice is uniformly driven.
A sketch of a spatiotemporally driven lattice for the case of three different driving laws ($n_p=3$) is shown in Fig 1. \\
In principle one could also imagine more complicated unit cells containing more barriers or even more complicated driving laws. The only restrictions that we have to make 
in order to employ the computational scheme as presented in the following section, is that the Hamiltonian remains periodic in time and that two neighbouring barriers do not 
have a notable overlap at any time. 

\begin{figure}[htbp]
\centering
\includegraphics[width=1.0\columnwidth]{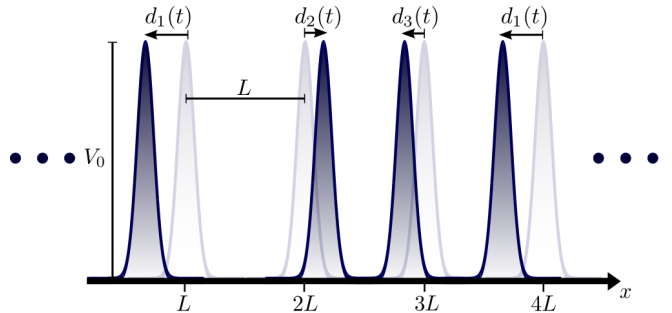}
\caption{\label{fig:lattice}Snapshot of a spatiotemporally driven lattice consisting of Gaussian barriers of height $V_0$ with lattice spacing $L$. Three different driving laws $d_i(t)$ for $i=1,2,3$ are
periodically repeated. The shaded barriers indicate the equidistant equilibrium positions of the barriers.  }
\end{figure}

\section{Computational scheme: the propagator method}
\label{S2}
In this section we develop a computational scheme in the framework of Floquet-Bloch theory 
which enables us to propagate an arbitrary initial state according to the Hamiltonian given by Eq. \ref{Hamiltonian}. 
Our formalism is based on the ideas presented in \cite{Shirley:1965} (see also \cite{Tannor}), and was originally designed to study atomic
and molecular multi-photon processes. In the following, we show how the formalism can be extended in order to describe the dynamics in spatiotemporally driven lattice as introduced in the previous section. 
In doing so, we have incorporated the possibility of a nonzero quasi-momentum as well as the complex nature of the unit cell which can contain several barriers, each equipped with a different driving law.

\subsection{Floquet-Bloch Theory}
To be self-contained, let us start by summing up the most important results from Floquet and Bloch theory.
Because the Hamiltonian under investigation is periodic in time with $H(x,t)=H(x,t+T)$ and $T=\frac{2\pi}{\omega}$, Floquets theorem ensures that every solution of Eq. \ref{Schrodinger} can be written as
\begin{equation}
 \Psi_{\alpha}(x,t)=e^{-i\epsilon_{\alpha}t} \Phi_{\alpha}(x,t),
\end{equation}
where the Floquet Mode (FM) $\Phi_{\alpha}$ respects the periodicity of the Hamiltonian, i.e. $\Phi_{\alpha}(x,t)=\Phi_{\alpha}(x,t+T)$ and $\epsilon_{\alpha}$ is a real number often termed the 'quasi-energy' (QE).
It is straightforward to see that adding or subtracting some integer multiple of $\omega$ to the QE while simultaneously multiplying the FM by an appropriate phase factor leaves the solution of the TDSE
$\Psi_{\alpha}(x,t)$ invariant. Hence, the QEs can always be chosen to be within the interval: $[-\frac{\omega}{2},\; \frac{\omega}{2}]$.\\
Knowing the FMs of our system is of particular relevance because it allows to compute the time evolution of any initial state. This is because they are 
eigenstates of the time evolution operator $U(t,t_0)$ over one period of the driving: 
\begin{equation}
U(T+t_0,t_0)\Phi_{\alpha}(x,t_0)=e^{-i\epsilon_{\alpha}T}\Phi_{\alpha}(x,t_0)
\label{eigenstates}
\end{equation}
 which follows directly from 
applying the time evolution operator $U(T+t_0,t_0)$ to the solution of the TDSE $\Psi_{\alpha}(x,t)$. Hence, the stroboscopic time evolution of an arbitrary initial state can be calculated as \cite{Schanz:2005}
\begin{equation}
 \Psi(x,mT+t_0)=\sum_{\alpha} C_{\alpha}(t_0) e^{-i\epsilon_{\alpha}mT} \Phi_{\alpha}(x,t_0).
  \label{time easy}
\end{equation}
where the $C_{\alpha}(t_0)$ are obtained as the overlap of the initial state with the FM $\Phi_{\alpha}(x,t_0)$.
So far we have only taken into account the temporal periodicity of our system.
However, the Hamiltonian considered in this work features spatial periodicity as well,
since $H(x,t)=H(x+n_p L,t)$. Accordingly, the FMs can be written in terms of Floquet-Bloch modes (FBMs) as $\Phi_{\alpha, \kappa}(x,t)=e^{i\kappa x} \phi_{\alpha, \kappa}(x,t)$ with  
$\phi_{\alpha, \kappa}(x,t)= \phi_{\alpha, \kappa}(x+ n_p L,t)= \phi_{\alpha, \kappa}(x,t+T)$ and $\kappa \in [-\pi/(n_p L),+\pi/(n_p L)]$ being the quasi-momentum. The stroboscopic time evolution for an initial state (Eq. \ref{time easy}) 
becomes \cite{Schanz:2005} then
\begin{equation}
\begin{aligned}
 \Psi(x,mT+t_0) &= \\
 \int_{-\pi / (n_p L)}^{+\pi / (n_p L)} &d\kappa \sum_{\alpha} C_{\alpha,\kappa}(t_0) e^{-i\epsilon_{\alpha,\kappa}mT } \Phi_{\alpha,\kappa}(x,t_0).
  \label{time}
\end{aligned}
\end{equation}

\subsection{Evaluation of the time evolution operator}
Apparently, once the FBMs $\Phi_{\alpha,\kappa}$ are known, any quantum state $\Psi(x,mT)$ can be propagated stroboscopically according to Eq. \ref{time}. In the following we explain how 
the FBMs can be obtained numerically in an efficient way.
The general idea is to make use of the fact that -as mentioned above- the FBMs are eigenstates of the one period time evolution operator $U(T+t_0,t_0)$. 
Accordingly, once the matrix representation of this operator is calculated in some basis 
it can be diagonalized and one obtains both the FBMs $\Phi_{\alpha,\kappa}(x,t_0)$ as well as the corresponding QEs $\epsilon_{\alpha, \kappa}$.

\subsubsection{The underlying Hilbert space} 
\label{hilbert space}
The first step towards the calculation of the matrix representation of the evolution operator, and with this of the FBMs,
is to specify the Hilbert space $\mathcal{H}$ in which the solutions of the TDSE (Eq.\eqref{Schrodinger}) can be represented. 
It was argued in \cite{Sambe:1973} that $\mathcal{H}$ can be composed as a product space of the Hilbert space of square integrable functions $\mathcal{R}$ and of the Hilbert space of time periodic functions $\mathcal{T}$.\\
In the following we define appropriate bases for $\mathcal{R}$,$\mathcal{T}$ and finally for $\mathcal{H}$. 
The states that we want to represent in $\mathcal{H}$, namely the FBMs, have to obey Bloch's theorem. This imposes that of the square integrable functions 
constituting the Hilbert space $\mathcal{R}$ we are only interested in the ones obeying $\Psi(x+n_pL)=e^{i\kappa n_p L} \Psi(x)$. We can take this into account by choosing quasi-momentum dependent 
basis vectors: $\braket{ x| \mu_{\kappa}} = \frac{1}{n_p L} e^{i(2\pi \mu /(L n_P) + \kappa)x} $ for $\mu \in \mathbb{Z}$ which are in accordance to the Bloch theorem since  
$\ket{ \mu_{\kappa}}=e^{i\kappa x}\ket{ \mu}$ where $\braket{x | \mu}=\braket{x + n_p L | \mu}$. 
Hence, once the time evolution operator is obtained in this basis and the FBMs are calculated as its eigenvectors, they will automatically satisfy Bloch's theorem because the basis 
in which they are expanded does.
The inner product in $\mathcal{R}$ can be defined as $\braket{f| g}=\int_{-\infty}^{\infty} dx f^*(x) g(x)$ for $\ket{f},\ket{g} \in \mathcal{R}$.
For the  Hilbert space of time periodic functions $\mathcal{T}$
the inner product is given by $\braket{a| b}=\frac{1}{T}\int dt \, a^*(t) b(t)$ for $\ket{a},\ket{b} \in \mathcal{T}$
and a natural choice for the basis vectors are the Fourier vectors  $\braket{ t| n}=e^{i n\omega t}$ with $n \in \mathbb{Z}$.\\
As mentioned earlier, the entire solution space $\mathcal{H}$ of Eq.\eqref{Schrodinger} can be constructed as the product space: $\mathcal{H}= \mathcal{R} \otimes \mathcal{T}$. Thus, 
a possible choice of basis vectors for $\mathcal{H}$ is given by the product basis: $\ket{\mu_{\kappa}} \otimes \ket{n} \equiv \ket{\mu_{\kappa},n} \rangle $ and the scalar product associated to $\mathcal{H}$ is:   
\begin{equation}
 \langle\braket{\Psi| \Psi'}\rangle= \frac{1}{T} \int_0^T  dt \int_{-\infty}^{+\infty} dx \Psi^*(x,t) \Psi'(x,t).
\end{equation}

\subsubsection{The time evolution operator} 
 
After we have specified the solution space of the TDSE we are ready to calculate its solutions, or more precisely its FBMs. As mentioned before, this can be done 
by finding the eigenvectors of the one period time evolution operator $U(T+t_0,t_0)$.\\  
It was shown in \cite{Shirley:1965} that the time evolution operator represented in our basis of $\mathcal{R}$ can be calculated for zero quasi-momentum as:
\begin{equation}
\begin{aligned}
  U_{\mu \nu}(t,t_0) &\equiv \bra{\mu} U(t,t_0) \ket{\nu} \\
  &=\sum_{n=-\infty}^{+\infty} \langle \bra{\mu n} e^{-iH_f(t-t_0)} \ket{\nu 0} \rangle e^{in\omega t},
\label{time evo operator no kap}
\end{aligned}
\end{equation}
where the Floquet operator $H_f(x,t)=H(x,t)-i \frac{\partial}{\partial t}$ was introduced. Because in this work the authors were interested in the 
time evolution of a single atom in a spatially homogeneous oscillating magnetic field, there was no need for the introduction of a nonzero quasi-momentum. 
We however, are interested in the action of a spatially periodic inhomogeneous potential and are therefore entitled to consider the case $\kappa \neq 0$ as well.
In order to do this, we exploit that states of different quasi-momenta are of course not mixed through the action of the Hamiltonian. Thus the matrix representation of the time evolution operator $U_{\mu \nu}(t,t_0)$
can be thought of being a block matrix, where every block acts only on states with a certain quasi-momentum $\kappa$. For each of these blocks we introduce the notation $U_{\mu \nu}^{\kappa}(t,t_0)$.  
Equation \eqref{time evo operator no kap} can be adjusted easily, by replacing the mere plane wave basis $\ket{\mu}$ with the in section \ref{hilbert space} introduced $\kappa$-dependent basis vectors $\ket{\mu_{\kappa}}$.
Thus we obtain for the evolution operator for now arbitrary quasi-momentum:
\begin{equation}
\begin{aligned}
 U_{\mu \nu}^{\kappa}(t,t_0) &\equiv \bra{\mu_{\kappa}} U(t,t_0) \ket{\nu_{\kappa}}\\
&=\sum_{n=-\infty}^{+\infty} \langle \bra{\mu_{\kappa} n} e^{-iH_f(t-t_0)} \ket{\nu_{\kappa} 0} \rangle e^{in\omega t},
\label{time evo operator}
\end{aligned}
\end{equation}
which is for $t=t_0+T$ the desired expression for the matrix elements of the time evolution operator over one driving period.  
In the following we show how Eq.\eqref{time evo operator} can be evaluated numerically.
To begin with, we divide the interval $(t_0,t_0+T)$ into $N$ small 
intervals of length $\Delta t$, thus allowing for a truncation of the exponential series for sufficiently small $\Delta t$. 
Hence, the quantity of interest becomes the evolution operator over the $j$th short time span $\Delta t$: $U_{\mu \nu}^{\kappa,j}=U_{\mu \nu}^{\kappa}(t_0+j\Delta t,t_0+(j-1)\Delta t)$ for $j=1,...,N$, while the 
full operator can be obtained as the product of all $U_{\mu \nu}^{\kappa,j}$ afterwards (we omit the matrix indices '$\mu$' and '$\nu$' for the sake of clarity):
\begin{equation}
 U^{\kappa}(t_0+T,t_0)=U^{\kappa,N}(t_0) \cdots  U^{\kappa,1}(t_0)=\prod_{j=1}^{N} U^{\kappa,j}(t_0).
\label{product}
\end{equation}
From Eq.\eqref{time evo operator} we get
\begin{equation}
\begin{aligned}
 U_{\mu \nu}^{\kappa,j} =& \lim\limits_{p_{\text{max}} \rightarrow \infty} \sum_{n=-\infty}^{+\infty} e^{in\omega j \Delta t}  \\
 \times  &\sum_{p=0}^{p=p_{\text{max}}} \frac{1}{p!} \left( -i \Delta t\right)^p \langle \bra{\mu_{\kappa} n} H_f^p(x,t-t_0) \ket{\nu_{\kappa} 0} \rangle ,
\end{aligned}
\label{time evo operator 2}
\end{equation}
where in a numerical calculation $p_{\text{max}}$ has to be sufficiently large to ensure convergence and  $\langle \bra{\mu_{\kappa} n} H_f^p(x,t) \ket{\nu_{\kappa} 0} \rangle$ are the matrix 
elements of the $p$-th power of the Floquet operator $H_f(x,t)$. 
In \cite{Shirley:1965} it was shown how these can be calculated recursively for $\kappa=0$. For nonzero quasi-momentum we obtain (see Appendix: \ref{A1})
\begin{equation}
\begin{aligned}
 \langle \bra{\mu_{\kappa} n} &H_f^p \ket{\nu_{\kappa} 0} \rangle \approx 
 \sum_{n'=-n_{\text{max}}}^{+n_{\text{max}}} \sum_{\mu'=-\mu_{\text{max}}}^{+\mu_{\text{max}}} \\
 &\times \left( H_{\mu \mu'}^{\kappa\, (n-n')} + n \omega \delta_{\mu \mu'}\delta_{n n'}\right) 
\langle \bra{\mu'_{\kappa} n'} H_f^{(p-1)} \ket{\nu_{\kappa} 0} \rangle.
\end{aligned}
\label{matrix floq}
\end{equation}
Thereby, $n_{\text{max}}$ and $\mu_{\text{max}}$ have to be sufficiently large to ensure convergence and
$H_{\mu \nu}^{\kappa\, (n-m)}$ is the matrix element of the $(n-m)$-th Fourier coefficient of $H(x,t)$ represented in the basis of $\mathcal{R}$, i.e.
\begin{equation}
\begin{aligned}
 H_{\mu \nu}^{\kappa\, (n-m)}&= \bra{\mu_{\kappa}} H^{(n-m)}(x) \ket{\nu_{\kappa}}\\
 &=\frac{1}{T} \int_0^T  dt  \bra{\mu_{\kappa}}  e^{-i\omega t (n-m)} H(x,t) \ket{\nu_{\kappa}}.
\end{aligned}
\end{equation}
It is straightforward to see that the Fourier components of the Hamiltonian are 
\begin{equation}
 H_{\mu \nu}^{\kappa , (n)} = \frac{1}{2}(\frac{2\pi}{n_p L} \mu + \kappa)^2 \delta_{\mu \nu}\delta_{n 0} + V_{\mu \nu}^{(n)}
\label{fourier hamiltonian}
\end{equation}
with $V_{\mu \nu}^{(n)}= \bra{\mu_{\kappa}} V^{(n)}(x) \ket{\nu_{\kappa}}$ where the $\kappa$-dependence can be omitted since 
the potential is a function solely of the position operator and not of its derivatives and therefore: 
$V_{\mu \nu}^{\kappa , (n)}=\bra{\mu_{\kappa}} V^{(n)}(x) \ket{\nu_{\kappa}}=\bra{\mu} e^{-i\kappa x} V^{(n)}(x) e^{i\kappa x}\ket{\nu}=\bra{\mu} V^{(n)}(x) \ket{\nu}=V_{\mu \nu}^{(n)}$. 

\subsubsection{The potential energy} 

So far we have seen how the time evolution operator $U_{\mu \nu}^{\kappa}(T+t_0,t_0)$ for a given quasi-momentum $\kappa$ can be calculated by evaluating the matrix elements of the powers of the Floquet operator 
as given by Eq.\eqref{matrix floq}, which in turn requires the calculation of the Fourier components of the Hamiltonian via Eq.\eqref{fourier hamiltonian}.
The remaining task is to compute the Fourier components of the potential $V^{(n)}(x)$, or more precisely their matrix representations in $\mathcal{R}$:  $V_{\mu \nu}^{(n)}$.
Note that the so far described formalism does not distinguish between uniform- and spatiotemporal driving. However, this becomes relevant for the Fourier decomposition of the potential 
$V^{(n)}(x)$ as we shall see in the following.\\ 
Let us start by considering the potential of a single oscillating barrier which will be denoted by $V_{\text{SB}}(x,t)$ in the following. 
Performing a Fourier transformation yields 
$V_{\text{SB}}(x,t)=\sum_{n=-\infty}^{+\infty} V_{\text{SB}}^{(n)}(x) e^{in\omega t}$
with $V_{\text{SB}}^{(n)}(x) \equiv \frac{1}{T}\int_0^T V_{\text{SB}}(x,t) e^{in\omega t}$.
If we now include a nonzero initial time $t_0$ as well as possible phase of the barrier motion of $\delta$ (cf. Eq. \eqref{driving}),
we get via the transformation $t \rightarrow t + t_0 + \frac{\delta}{\omega}$:
\begin{equation}
  V_{\text{SB}}(x,t)=\sum_{n=-\infty}^{+\infty} V_{\text{SB}}^{(n)}(x) e^{in(\omega(t+t_0) + \delta)}
\end{equation}
Apparently, the phase shift $\delta$ of a barrier leads to a complex phase factor of $e^{in \delta}$ of the corresponding Fourier coefficient.
For $n_p$ barriers each with a different phase $\delta_i$ this generalizes to 
\begin{equation}
 V^{(n)}(x)=\sum_{i=1}^{n_p} V_{\text{SB}}^{(n)}(x-x_{0,i})e^{in (\omega t_0 +\delta_i)},
\label{fourier coeff}
\end{equation}
with $x_{0,i}=iL$ being the equilibrium position of the $i$th barrier.
For the desired matrix representation of this Fourier mode $V_{\mu \nu}^{ (n)}$ this yield (see Appendix \ref{A2})
\begin{equation}
 V_{\mu \nu}^{(n)}=\sum_{i=1}^{n_p} V_{\text{SB},\, \mu \nu}^{(n)}  e^{i(n (\omega t_0 +\delta_i)+ \frac{2\pi}{Ln_p}(\mu -\nu)x_{0,i})},
\end{equation}
where $V_{\text{SB},\, \mu \nu}^{ (n)}=\bra{\mu} V_{\text{SB}}^{(n)}(x) \ket{\nu}$ is the $n$-th Fourier component of a single oscillating barrier 
represented in our basis of $\mathcal{R}$.\\
At this point we have all the ingredients to make use of Eq.\ref{time evo operator 2} in order to determine the one-period time evolution operator and thus the FBMs, the QEs and finally 
the stroboscopic time evolution of arbitrary initial states via Eq.\eqref{time}. 

\subsection{Computation for different initial times}

The final remark of this section shall be on the role of the initial time within the described formalism. Due to the linearity of the Schr\"odinger equation, the asymptotic behaviour of an observable 
in a time-dependent system depends, in general, on the initial time $t_0$. E.g. in the context of ratchet physics it was shown that the asymptotic transport velocity of an initial state depends crucially on $t_0$ 
(see \cite{Salger:2009} and also section \ref{S5}).
Thus, in numerical simulations the time propagation typically has to be performed for many different initial times in order to capture the full physical behaviour. 
The straightforward way to include different initial times in the formalism as described above, is to simply plug in the potential $V(x,t+t_0)$ into 
the calculation of the Fourier components of the Hamiltonian (via Eq. \eqref{fourier hamiltonian}).   
The downside of this approach is that the entire formalism to calculate $U(T+t_0,t_0)$ has to be carried out for each considered value of $t_0$, which can be quite time consuming.
Fortunately, there is a much faster way. Within the presented formalism we have calculated the time evolution operator over an entire driving period as the product 
of $N$ operators $U^{\kappa,j}(t_0)$ where each of the $U^{\kappa,j}(t_0)$ propagates the small time step $\Delta t$ (see . Eq.\eqref{product}). The idea is now to calculate the $N$ operators
$U^{\kappa,j}(t_0)$ for some value of the initial time, say for $t_0=0$. The desired operator $U(T+t_0,t_0)$ for arbitrary $t_0$ can be obtained simply as the product of all the $U^{\kappa,j}(t_0=0)$, where the 
dependence on $t_0$ is now captured in the
ordering of the operators. More precisely, Eq.\eqref{product} can be rewritten for $0 \leq t_0 \leq T $ as:
\begin{equation}
\begin{aligned}
 U^{\kappa}(t_0+T,t_0) &= \left( U^{\kappa,j_0-1} \cdots  U^{\kappa,1}    \right) \cdot  \left( U^{\kappa,N} \cdots  U^{\kappa,j_0}    \right)\\
      &= \prod_{j=1}^{j_0-1} U^{\kappa,j}(t_0=0) \cdot   \prod_{j=j_0}^{N} U^{\kappa,j}(t_0=0)
\label{product2}
\end{aligned}
\end{equation}
with $j_0=\lceil N/T \cdot t_{0}\rceil$ where $\lceil x\rceil$ denotes the smallest integer number larger than $x$. The obvious advantage is, that the $U^{j,\kappa}(t_0)$ have to be calculated only 
for one initial time. Afterwards, the time evolution operators for arbitrary initial times can be calculated easily by means of matrix multiplication. 
For the calculation of the FBMs in the setup of the spatiotemporally driven lattice, the speed up of 
this procedure to include the dependence of the initial time -as compared to the previously mentioned straightforward way-
proved to be significant and amounted to up to one order of magnitude.

\section{Symmetry analysis}
\label{S3}
It goes without saying that classifying the symmetries of a physical system is often helpful in order to understand the phenomena occurring in it. 
In the context of quantum ratchets for example it was extensively studied how the symmetries of the Floquet operator affect the possibility of directed particle motion \cite{Denisov:2007, Hanggi:2014}.
Within this section we present the relevant symmetries of the Hamiltonian and deduce their consequences on the time evolution operator as well as on the FBMs.
In doing so we start with time reversal- and parity symmetry which are commonly studied in the context of uniformly driven 1D lattices. Afterwards, we investigate the impact of the spatiotemporal driving.

\subsection{Time reversal symmetry}

The Hamiltonian is symmetric under time reversal if $H(x,t)=H(x,-t+\tau)$ for some appropriate time shift $\tau$, which we will assume to be zero in the following. 
Let us start with investigating the consequences of time reversal symmetry on the time evolution operator $U(t,t_0)$. 
For the simpler case of
zero quasi-momentum it was shown in \cite{Graham:1991} that the matrix elements of $U(T,0)$ in the plane wave basis as used in this work obey:
\begin{equation}
\begin{aligned}
U_{-\nu -\mu}\left(T/2,0\right)&=U_{\mu \nu}\left(T, T/2\right), \\
U_{-\nu -\mu}(T,0)&=U_{\mu \nu}(T,0)
\end{aligned}
\label{symmetries}
\end{equation}
Of particular interest is the last named symmetry because it concerns the time evolution operator over an entire driving period, which is the one used to 
determine the FBMs (see Eq.\ref{eigenstates}). By employing ideas from section \ref{S2} one can readily generalize this symmetry to 
nonzero values of the quasi-momentum $\kappa$:  
At the heart of the calculation of the matrix elements of the time evolution operator as given by Eq.\eqref{time evo operator} 
is the calculation of the Fourier components of the Hamiltonian as given by Eq.\eqref{fourier hamiltonian}. Note that the quasi-momentum $\kappa$ 
only enters in the diagonal term which is proportional to $(\frac{2\pi}{n_p L} \mu + \kappa)^2 \delta_{\mu \nu}$. Apparently, this term is invariant under $\mu \rightarrow -\nu$ 
iff we set  $\kappa \rightarrow -\kappa$ simultaneously. Thus the symmetry generalizes for arbitrary quasi-momentum $\kappa$ to:
\begin{equation}
 U_{-\nu -\mu}^{-\kappa}(T,0)=U_{\mu \nu}^{\kappa}(T,0).
\end{equation}
In fact we find strong evidence that both symmetries as stated in Eq.\eqref{symmetries} are special cases of the more general symmetry which holds here:
\begin{equation}
 U_{-\nu -\mu}^{-\kappa}(t_2,t_1)=U_{\mu \nu}^{\kappa}(T-t_1,T-t_2)
 \label{new symmetry}
\end{equation} 
with $0<t_1<t_2<T$. In appendix \ref{A3} we provide a rigorous proof 
up to first order in the expansion of the time evolution operator, i.e. for 
$p_{\text{max}}=1$ in Eq.\eqref{time evo operator 2}. Beyond first order, we have strong numerical evidence for the validity of Eq.\eqref{new symmetry}.\\
Finally, we consider the consequence of a time reversal symmetry of the Hamiltonian on the FBMs. It was argued in \cite{Hanggi:2014} that these must obey
\begin{equation}
 \Phi_{\alpha,\kappa}(x,t)=\sigma_{\alpha}  \Phi^*_{\alpha,-\kappa}(x,T-t), \quad \sigma_{\alpha}=\pm 1.
\end{equation}
For the representation as chosen in this work, this yields for the components of the FBMs 
\begin{equation}
 \braket{\mu | \Phi_{\alpha,\kappa}(x,t)} \equiv \Phi^{\mu}_{\alpha,\kappa}(t) = \sigma_{\alpha} (\Phi^{-\mu}_{\alpha,-\kappa}(T-t))^*.
\label{comps t}
\end{equation}

\subsection{Parity symmetry}

The Hamiltonian is said to be invariant under parity symmetry if $H(x,t)=H(-x+\chi,t+T/2)$ for some appropriate spatial shift $\chi$ which we can assume without loss of generality to be zero. 
As argued in \cite{Graham:1991} parity symmetry yields for the time evolution operator:
\begin{equation}
 U_{\mu \nu}^{\kappa}(T,0)= \sum_{\theta} U_{-\mu -\theta}^{\kappa}(T/2,0)  U_{\theta \nu}^{\kappa}(T/2,0)
\end{equation}
This relation can be of particular use since it allows us to half the computational effort. 
Furthermore, this symmetry of the time evolution operator leads to a symmetry of the FBMs \cite{Hanggi:2014}:
\begin{equation}
 \Phi_{\alpha,\kappa}(x,t)=\sigma_{\alpha}  \Phi_{\alpha,-\kappa}(-x,t+T/2), \quad \sigma_{\alpha}=\pm 1.
\end{equation}
In analogy to the time reversal symmetry (see Eq.\eqref{comps t}) this yields for the components of the FBMs in the representation introduced in section \ref{S2}:
\begin{equation}
\Phi^{\mu}_{\alpha,\kappa}(t) = \sigma_{\alpha} \Phi^{-\mu}_{\alpha,-\kappa}(t+T/2).
\label{comps x}
\end{equation}

\subsection{Parity and time reversal symmetry}

Apparently, there is the possibility for a Hamiltonian to be symmetric under both parity and time reversal symmetry. In this case it obeys $H(x,t)=H(-x,-t+\tau)$. 
The components of the Floquet modes have to fulfill both Eq.\eqref{comps t} as well as Eq.\eqref{comps x}. Hence we get:
 \begin{equation}
\Phi^{\mu}_{\alpha,\kappa}(T-t) = \sigma_{\alpha} (\Phi^{\mu}_{\alpha,\kappa}(t+T/2))^*.
\label{comps xt}
\end{equation}

\subsection{Shift symmetry}
\label{shift symmetry}
Finally we turn our focus to the impact of spatiotemporal driving and investigate its consequences on both the time evolution operator as well as on the FBMs. 
Let us as an introductory example consider a lattice with a unit cell which contains three barriers, i.e. we have $n_p=3$. We choose for the three different initial phases $\delta_i$ for $i=1,2,3$
in the driving law (see Eq.\eqref{driving}) $(\delta_1=0,\delta_2, \delta_3=0)$ with $\delta_2 \in [0,2\pi)$. Hence, we obtain a driven lattice where the central barrier of each unit cell is potentially
out of phase compared to its two neighbouring barriers. In the following we try to deduce properties of the overall structure of the time evolution operator for such a setup.
In doing so, we will show that this 'partial shift symmetry breaking' induced by the complex nature of the unit cell
in a spatiotemporally driven lattice has profound consequence both on the time evolution operator as 
well as on the FBMs, which have to the best of our knowledge not yet been discussed in the literature.\\  

\begin{figure}[htbp]
\centering
\includegraphics[width=1.0\columnwidth]{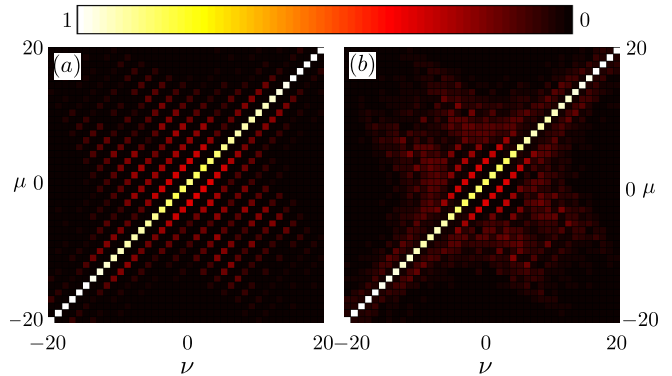}
\caption{\label{fig:time_evo_op} Absolute value of matrix elements $U_{\mu \nu}(T,0)$ of the time evolution operator for $\kappa=0$. The phases $\delta_i$ of the three barriers within one unit cell 
are $(0,0,0)$ in a) and $(0,\pi,0)$ in b). The remaining parameters are: $L=10, \ V_0=1.0, \ \omega=1.0, A=1.0$ and $\Delta=0.5$.}
\end{figure}

To get some insight we calculate numerically the matrix elements of the time evolution operator $U_{\mu \nu}(T,0)$ for zero quasi-momentum and for $t_0=0$. We do this for the case of a nonzero phase shift of the central barrier
of $\delta_2=\pi$ and compare the results to the uniformly driven lattice with $\delta_2=0$. The absolute values of the obtained $U_{\mu \nu}(T,0)$ are shown in Fig. \ref{fig:time_evo_op}. 
The most intriguing feature is that in both cases we observe a stripe like structure for the nonzero elements. 
In particular, for the uniformly driven case (Fig. \ref{fig:time_evo_op} (a)) we see that for a fixed value of $\mu$, only every third value of $\nu$ corresponds to a nonzero 
matrix element and vice versa. 
For the spatiotemporally driven lattice (Fig. \ref{fig:time_evo_op} (b)) this behaviour persists for elements close to the main diagonal but becomes less pronounced further away from it, i.e. for 
larger values of $|\mu-\nu|$.\\
In order to understand the overall structure of the two shown time evolution operators we consider their different symmetries under spatial shifts. Obviously both Hamiltonians obey $H(x,t)=H(x+n_pL,t)$
because by construction the length of the unit cell was chosen to be $n_pL$. However, in the uniformly driven case, the Hamiltonian additionally obeys $H(x,t)=H(x+L,t)$ and consequently 
the time evolution operator $U_{\mu \nu}(T,0)$ should commute with the operator $S^{L}=e^{-iL \hat p }=e^{-L\frac{\partial}{\partial x}}$ performing a spatial shift of $L$. 
This is because the FBMs form a complete set and are eigenstates of both $U_{\mu \nu}(T,0)$ and $S^{L}$.
For arbitrary quasi-momentum $\kappa$ the matrix representation 
of the shift operator $S^{L}$ becomes
\begin{equation}
 S^{\kappa,L}_{\mu \nu}= \bra{\mu_{\kappa}} S^L \ket{\nu_{\kappa}}= e^{-i(\frac{2 \pi}{n_p L}\mu + \kappa )L } \delta_{\mu \nu}.
\end{equation}
Thus the requirement of commutation with $S^{\kappa,L}_{\mu \nu}$ yields for the matrix elements $U_{\mu \nu}(T,0)$ (we omit the argument for the sake of clarity):
\begin{equation}
   S^{\kappa,L}_{\mu \beta} U_{\beta \nu} - U_{\mu \gamma} S^{\kappa,L}_{\gamma \nu}=\left( e^{-i\frac{2 \pi}{n_p}\mu} - e^{-i\frac{2 \pi}{n_p}\nu}  \right) U_{\mu \nu} \overset{!}{=} 0
\end{equation}
Apparently, this requires that either $U_{\mu \nu}(T,0)=0$ or $\mu-\nu=n_p z$ for $z \in \mathbb{Z}$ and explains the stripe like structure of the time evolution operator as observed in 
Fig. \ref{fig:time_evo_op} (a). Arguments along a very similar line lead to a restriction on the components of the FBMs $\Psi^{\mu}_{\alpha,\kappa}(t)$. Due to the Bloch theorem the FBMs must be eigenstates of 
$S^{L}$ and thus:
\begin{equation}
  S^{\kappa,L}_{\mu \nu} \Psi^{\nu}_{\alpha,\kappa}(t)=e^{-i\frac{2 \pi}{n_p}\mu } e^{-i \kappa L} \Psi^{\mu}_{\alpha,\kappa}(t) \overset{!}{=} \lambda_{\alpha}  \Psi^{\mu}_{\alpha,\kappa}(t) 
\label{eigenstates uniform}
\end{equation}
for some complex eigenvalue $\lambda_{\alpha}$. This requires that the prefactor $e^{-i\frac{2 \pi}{n_p}\mu }$ must be independent of $\mu$ which is only true if all the nonzero 
components of the FBM $\Psi^{\mu}_{\alpha,\kappa}(t)$ can be labeled by $\mu=n_p z + q$ for $z \in \mathbb{Z}$ and $q=0,1,...,n_p-1$. The corresponding $n_p$ different eigenvalues are
$\lambda_{\alpha, q}=e^{-i(\frac{2 \pi}{n_p}q + L \kappa)}$ .\\
Although the reported restrictions on the matrix elements $U_{\mu \nu}$ as well as on the FBMs were derived for the uniformly driven lattice with $\delta_2=0$, we see clearly
that, for the case of the evolution operator, this structure survives to some degree even for the largest possible phase shift of $\delta_2=\pi$ (cf. Fig.  \ref{fig:time_evo_op} (b)).
For decreasing $\delta_2$ we observe that the uniformly driven case is approached closer and closer.

\begin{figure}[htbp]
\centering
\includegraphics[width=1.0\columnwidth]{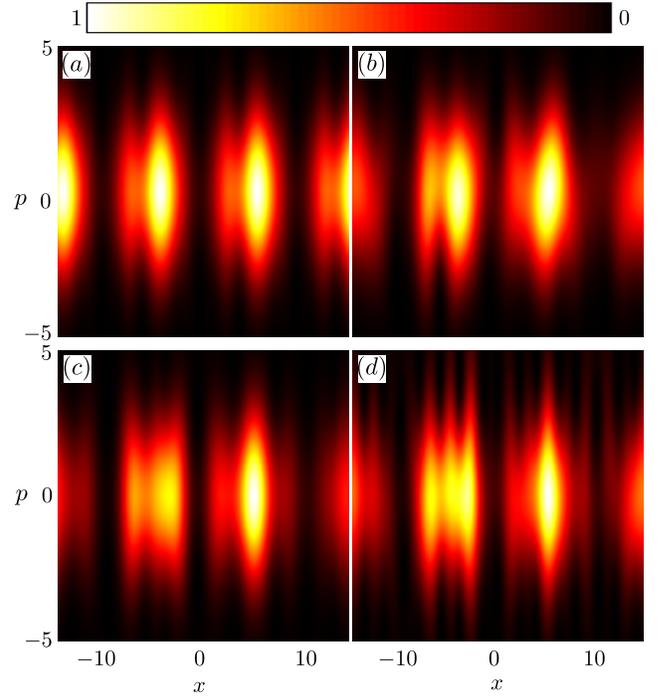}
\caption{\label{fig:husimis} Husimi distributions in arbitrary units for $\sigma=0.5$ of a FBM at zero quasi-momentum in a lattice with $n_p=3$ for 4 different settings of the barrier phases  $(\delta_1,\delta_2,\delta_3)$:
(a) $(0,0,0)$, (b) $(0,\pi/2,0)$, (c) $(0,\pi,0)$, (d) $(0,\pi,\pi/4)$. The barriers equilibrium positions are at $x_1=-10$, $x_2=0$ and $x_3=10$. Remaining parameters as in Fig. \ref{fig:time_evo_op}.}
\end{figure}

\subsection{Husimi representations for uniform and spatiotemporal driving}

We conclude the section on the symmetry analysis by analyzing the consequences of the discussed symmetries on Husimi representations,
which are a very commonly used tool to obtain a coarse grained visualization of a quantum state \cite{Takahashi:1989}.\\
The Husimi representation of a quantum state is defined by the square of the absolute value of its overlap with a coherent state $\ket{\rho(x,p)}$ centered around
position $x$, momentum $p$ and with width $\sigma$ \cite{Takahashi:1989}. Such a coherent state can be expressed as:
$\braket{\tilde{x}|\rho(x,p)} =(\pi \sigma^2)^{-1/4} e^{-\frac{(x-\tilde{x})^2}{2\sigma^2}  + ip\tilde{x} }$. 
For a FBM $\Phi_{\kappa}(x,t)$ with quasi-momentum $\kappa$ we can calculate the 
Husimi distribution $Q_{\kappa}(x,p,t)$ as:
\begin{equation}
\begin{aligned}
 Q_{\kappa}(&x,p,t)  =\frac{1}{2\pi}|\braket{\rho(x,p)|\Phi_{\kappa}}|^2\\
	  &=\frac{(\pi \sigma^2)^{-\frac{1}{4}}}{2\pi} \left| \int d \tilde{x} \; e^{-\frac{(x-\tilde{x})^2}{2\sigma^2} + ip\tilde{x}} \Phi_{\kappa}(\tilde{x},t) \right|^2\\
	   &= \sqrt{2\pi}  \left| \sum_{\mu} \Phi^{\mu}_{\kappa} (t) \; e^{-\frac{\sigma^2}{2}(\frac{2 \pi}{n_pL}\mu+\kappa-p)^2 + i(\frac{2 \pi}{n_pL}\mu + \kappa -p)x} \right|^2
\label{husimi calculate}
\end{aligned}
\end{equation}
where we have plugged in the expansion of the FBM $\Phi_{\kappa}(x,t)$ in terms of our basis of $\mathcal{R}$: $\Phi_{\kappa}(x,t)=\sum_{\mu} \Phi^{\mu}_{\kappa} (t) e^{-i (\frac{2 \pi}{n_pL}\mu+\kappa) x}$.
To get some insight, let us consider the Husimi representation of one specific FBM. This mode , denoted as $\Phi_G(x,t)$, has zero quasi-momentum and
is characterized as the FBM with the largest overlap with a spatially uniform state. It is of particular importance since it usually also has the largest overlap with a quantum particle which is 
initially distributed over many lattice sites- a situation that is commonly considered both in theory as well as in experiments \cite{Denisov:2007}. 
The Husimi distributions
$Q(x,p,t)$ of $\Phi_G(x,t)$ for $t=0$ are shown for different setups each containing three barriers per unit cell ($n_p=3$) but 
with different settings of the barrier phases $(\delta_1,\delta_2,\delta_3)$ (cf. Eq.\eqref{driving}) in Fig. \ref{fig:husimis}.
For the uniformly driven lattice (Fig. \ref{fig:husimis} (a)) we observe that the Husimi representation is invariant under a spatial shift of one barrier distance $L$, i.e. $Q(x,p,0)=Q(x+L,p,0)$. 
Apparently, this shift symmetry is broken for setups with nonzero barrier phases (Figs. \ref{fig:husimis} (b),(c),(d)). Even more, for the setups with $(0,0,0)$ and $(0,\pi,0)$ the Husimi 
distribution is symmetric with respect to an inversion of momentum, i.e. we observe $Q(x,p,0)=Q(x,-p,0)$. In the following we argue how these symmetries of 
the Husimi representations can be deduced from the corresponding symmetry analysis. \\ 
Let us start with the observed shift symmetry $Q(x,p,0)=Q(x+L,p,0)$. In fact, 
it follows directly from the shift symmetry of the FBM which obeys $\Phi_G(x,t)=\Phi_G(x+L,t)$ (cf. section \ref{shift symmetry}) that the shift symmetry of the 
Husimi distribution indeed holds for all times. 
Likewise, the observed symmetry of $Q(x,p,0)=Q(x,-p,0)$ can be understood conveniently with the help of the above symmetry analysis. 
By virtue of Eq.\eqref{comps t} we know that $\Phi^{\mu}_G (0)= (\Phi^{-\mu}_G (0))^*$ which can be shown easily to imply $Q(x,p,0)=Q(x,-p,0)$. More generally, one can show from Eq.\eqref{husimi calculate} that
the restrictions on the FBMs in the presence of time reversal symmetry (see Eq.\eqref{comps t}) yield for the Husimi representation:
\begin{equation}
 Q^{\kappa}(x,p,t)=Q^{-\kappa}(x,-p,T-t).
\label{husimi t}
\end{equation}
Analogously, the presence of parity symmetry, where the components of the FBMs obey Eq.\eqref{comps x}, implies for the Husimi representation:
\begin{equation}
 Q^{\kappa}(x,p,t)=Q^{-\kappa}(-x,-p,t+T/2).
\label{husimi x}
\end{equation}   
Note that even though we used a particular representation of the FBMs in order to derive the two latter relations for the Husimi distribution,
these relations themselves must of course hold for every other representation too. Thus Eqs. \eqref{husimi t} and \eqref{husimi x} are general results for FBMs in 1D driven lattices with time reversal- or
parity symmetry.\\
Because the Hamiltonians underlying Figs. \ref{fig:husimis} (a) and (c) with phases $(0,0,0)$ and $(0,\pi,0)$ respect both time reversal- as well as parity symmetry, the Husimi distributions 
obey Eq.\eqref{husimi t} and Eq.\eqref{husimi x}. The setup with $(0,\pi/2,0)$ as used in Fig. \ref{fig:husimis} (b) only inherits parity symmetry, i.e. the associated $Q(x,p,t)$
only obeys  Eq.\eqref{husimi x}. Finally, for $(0,\pi,\pi/4)$ no symmetry remains.    

\begin{figure}[htbp]
\centering
\includegraphics[width=1.0\columnwidth]{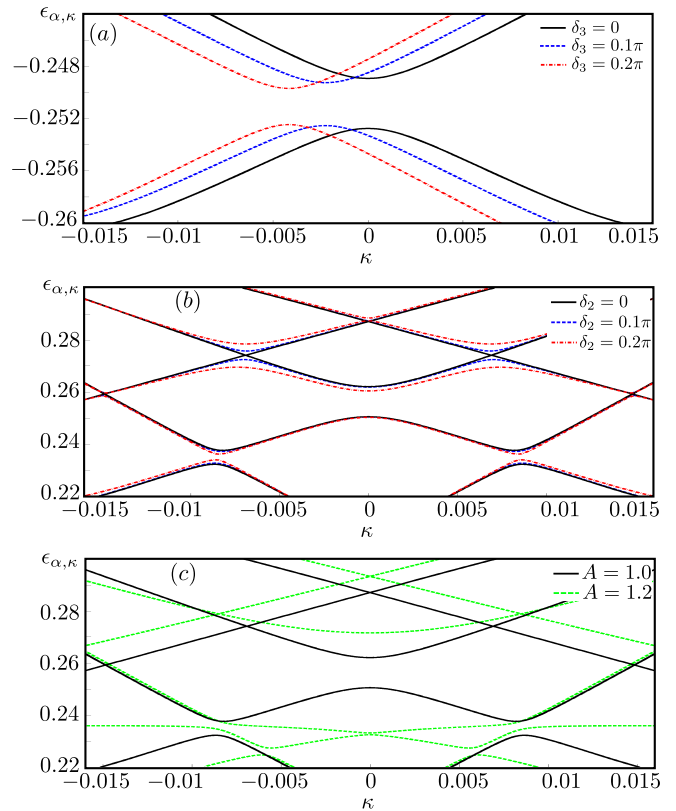}
\caption{\label{fig:spectrum} Extracts of the Floquet spectra for different setups containing three barriers per unit cell. The three phases of the barriers driving laws are (a) $(0,2\pi/3,\delta_3)$, 
(b) $(0,\delta_2,0)$ and (c) $(0,0,0)$. In (c) the spectra for two different global driving amplitudes are shown. Remaining parameters as in Fig. \ref{fig:time_evo_op}.  }
\end{figure}

\section{The Floquet spectrum}    
\label{S4}

The quasi-energies $\epsilon_{\alpha,\kappa}$ of the FBMs $\Psi_{\alpha,\kappa}(x,t)$ evaluated as functions of the quasi-momentum $\kappa$ constitute the quasi-energy- or Floquet spectrum of a periodically 
driven system. In the following we investigate 
the impact of the symmetries as introduced in section \ref{S3}
on the Floquet spectrum. This question has been subject of comprehensive research in the 
case of uniform driving (see e.g. \cite{Schanz:2005, Flach:2007, Denisov:2007, Hanggi:2014}), and we are going to sum up the most important results. 
However, our main interest is on the impact of the partially broken shift symmetry induced by the spatiotemporal driving.
As before we consider as an exemplary setup a lattice with a unit cell containing three barriers with phases: $(\delta_1,\delta_2,\delta_3)$.\\

\subsection{Impact of parity- and/or time reversal symmetry}

An extract of the Floquet spectrum for a phase configuration of $(\delta_1=0,\delta_2=\frac{2\pi}{3},\delta_3=0)$ is shown in Fig. \ref{fig:spectrum} (a). In this case 
the Hamiltonian is invariant under parity symmetry as introduced in  section \ref{S3} and
we observe that $\epsilon_{\alpha,\kappa}=\epsilon_{\alpha,-\kappa}$, i.e. the spectrum is symmetric with respect to $\kappa=0$. In fact it was argued in   
\cite{Hanggi:2014} that the invariance of the Hamiltonian under either parity- or time reversal symmetry generally yields a spectrum which is symmetric with 
respect to $\kappa=0$. In accordance to this, it is shown how a breaking of these symmetries by setting a second barrier phase, in this case $\delta_3$, to a nonzero value leads to a desymmetrization of the spectrum. 
Note that for such a configuration where we fix two barrier phases to $\delta_1=0$ and arbitrary $\delta_2$, parity symmetry is always present for $\delta_3=0$. Thus by deviating the value of
the phase of the third barrier $\delta_3$ from zero, one can very reliably tune the asymmetry of the spectrum.       

\subsection{The role of shift symmetry and exact- to avoided crossings transitions}

In the following we show how the deviation from a uniform driving towards a spatiotemporal driving affects the Floquet spectrum in a unique way. As a model system we consider a phase configuration 
of the barriers in a unit cell of  $(0,\delta_2,0)$, where as in section \ref{shift symmetry} the shift symmetry is said to be 'partially broken' for $\delta_2 \neq 0$. 
Representative extracts of the spectrum for uniform driving ($\delta_2 =0$) as well as for a spatiotemporal driving ($\delta_2 \neq 0$) are shown in Fig. \ref{fig:spectrum} (b). The most notable 
effect is, that the crossings which are exact ones for the uniform driving are casted into avoided crossings in the case of the spatiotemporal driving. At the same time the crossings which are avoided ones 
for uniform driving, remain avoided crossings for $\delta_2 \neq 0$. 
Even more, we see that at least to a certain degree one can control the width of the avoided crossing by tuning the phase of the central barrier $\delta_2$. 
This is best seen for the two crossings at $\epsilon \approx 0.275$ where the widths can be seen to increase for increasing $\delta_2$.
At this point it is worth mentioning that the widths of avoided crossings in Floquet spectra are crucial for many nonequilibrium phenomena reported in driven lattice setups,
such as Landau-Zener transitions \cite{Flach:2007}, the diffusion properties of a wave-packet \cite{Graham:1994} or the occurrence of an absolute negative mobility \cite{Weitz:2013}.
Hence, the additional flexibility introduced by the site-dependent driving should contribute to an increased controllability of the aforementioned effects.
For comparison we also show the spectrum of a uniformly driven lattice 
for two different driving amplitudes $A=1.0$ and $A=1.25$ in Fig. \ref{fig:spectrum} (c). Apparently, even though the spectrum is altered significantly by the increased driving amplitude, the 
nature of the crossings, i.e. exact or avoided, is not changed. \\
By means of the symmetry analysis carried out in section \ref{shift symmetry} it is quite straightforward to understand why the variation in the barrier phase $\delta_2$ 
leads to a transformation of the crossings from exact to avoided while the variation in the global driving amplitude $A$ did not. The key observation lies within 
Eq.\eqref{eigenstates uniform}, stating that the FBMs must be eigenstates of the operator performing a spatial shift of the barrier distance $L$. As argued, the allowed 
eigenvalues for a FBM $\Phi_{\alpha,\kappa}(x,t)$ are then given by $\lambda_{\alpha, q}=e^{-i(\frac{2 \pi}{n_p}q + L \kappa)}$ for $q=0,1,...,n_p-1$. Thus the FBMs can be separated 
into $n_p$ different symmetry classes, each characterized by one of the $n_p$ different eigenvalues $\lambda_{\alpha, q}$. According to the non-crossing rule \cite{Neumann:1929} 
states belonging to different symmetry classes are allowed to cross, while states within the same symmetry class cannot. 
For a nonzero value of $\delta_2$ the shift symmetry $x\rightarrow x+L$ gets destroyed and Eq.\eqref{eigenstates uniform} becomes invalid. Hence, the FBMs cannot be separated into different symmetry classes associated
to the shift operator $S^L$ and thus they are not allowed to cross anymore. At this point it is important to note that, even for a partially broken shift symmetry, the spectrum may still feature exact crossings 
due to the presence of different symmetry classes associated to 
parity- or time reversal symmetry.\\

\section{Directed transport}
\label{S5}

In this section, we study the possibility of an asymptotic particle current in the setup of a spatiotemporally driven lattice. 
The appearance of such currents in the absence of any mean forces has been studied intensively over the last two decades 
for uniformly driven lattices (see for example \cite{Hanggi:2005, Hanggi:2009, Hanggi:2014} and references therein). 
Similar to the previous sections, we begin with some general considerations and investigate the spatiotemporally driven lattice in particular afterwards.\\

\begin{figure}[htbp]
\centering
\includegraphics[width=1.0\columnwidth]{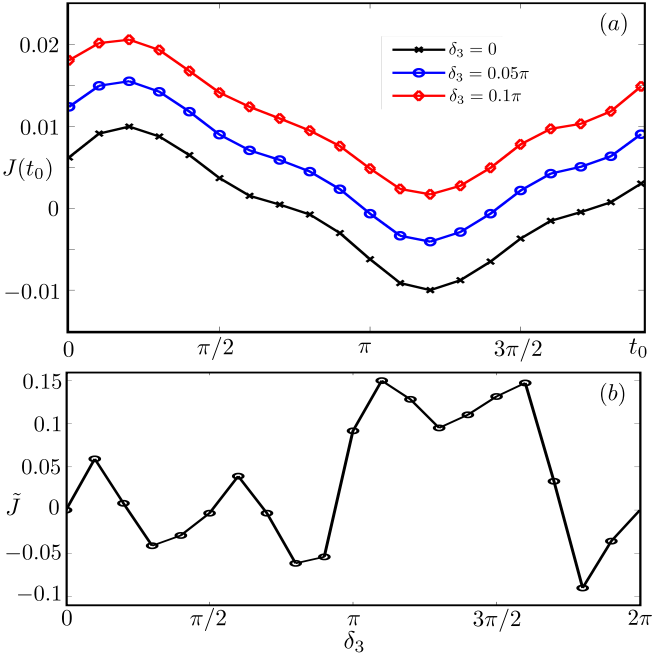}
\caption{\label{fig:transport} (a) Asymptotic quantum current $J(t_0)$ for three different configurations of the barrier phases: $(0,2\pi/3,\delta_3)$.
(b) Time averaged quantum current as function of the third barrier phase $\delta_3$. The lines are guides for the eye.
Remaining parameters as in Fig. \ref{fig:time_evo_op}.}
\end{figure}

\subsection{Asymptotic currents}

Following \cite{Denisov:2007} we define the asymptotic quantum current as the time averaged expectation value of the momentum operator for some initial state  $\ket{\Psi_I(t_0)}$:
\begin{equation}
 J(t_0)= \lim\limits_{t \rightarrow \infty} \frac{1}{t-t_0} \int_{t_0}^t d \tilde{t} \; \braket{ \Psi_I(\tilde{t})| \hat{p} |  \Psi_I(\tilde{t})}.
\end{equation}
One of the perks of using Floquet theory in order to study the dynamics of a time-dependent system is that once the FBMs are known, the asymptotic current 
can be calculated very conveniently as \cite{Denisov:2007, Schanz:2005}:
\begin{equation}
 J(t_0)=\int d\kappa \sum_{\alpha} v_{\alpha \kappa}  |C_{\alpha \kappa} (t_0)|^2  
\label{currents1}
\end{equation} 
where $v_{\alpha \kappa}$ is the averaged momentum of the FBM $\ket{\Psi_{\alpha \kappa}}$ and $C_{\alpha \kappa} (t_0)$ 
is the overlap of the initial state with the FBM $\ket{\Phi_{\alpha \kappa}}$ at time $t=t_0$. 
For a commonly studied initial state of a Gaussian wave packet: $\Psi_{I}(x,t_0)=(\pi \sigma^2)^{-1/4} e^{-\frac{x^2}{2\sigma^2}}$ these two quantities
can be calculated as:  
\begin{equation}
\begin{aligned}
 v_{\alpha \kappa}&=\frac{1}{T} \int_0^T dt \braket{\Phi_{\alpha \kappa}|\hat p |\Phi_{\alpha \kappa}} \\
    &= \frac{1}{T} \sum_{\mu}  \int_0^T dt  \left(\frac{2\pi}{L n_p} \mu + \kappa \right) |\Phi^{\mu}_{\alpha \kappa} (t)|^2 \\  
 C_{\alpha \kappa} (t_0) &=\braket{\Psi_{I} |  \Phi_{\alpha \kappa}}\\
    &=\frac{\sqrt{2\sigma}}{\pi^{-1/4}} \sum_{\mu} e^{-\frac{\sigma^2}{2L^2 n_p^2} (n_p L \kappa + 2\pi \mu)^2 } \Phi^{\mu}_{\alpha \kappa} (t_0)
\label{currents2}
\end{aligned}
\end{equation}
where as before the $\Phi^{\mu}_{\alpha \kappa} (t)$ are the components of the FBMs in our chosen basis which can be calculated according to the numerical scheme presented in section \ref{S2}.

\subsection{Symmetries and directed transport}
Now that we have seen how the asymptotic current $J(t_0)$ can be expressed through the components of the FBMs $\Phi^{\mu}_{\alpha \kappa} (t)$ we are able to 
analyze how the symmetries derived in section \ref{S3} for the $\Phi^{\mu}_{\alpha \kappa} (t_0)$ are carried over into symmetries of $J(t_0)$.\\
For a time reversal symmetric Hamiltonian we had seen that $\Phi^{\mu}_{\alpha,\kappa}(t) = \pm (\Phi^{-\mu}_{\alpha,-\kappa}(T-t))^*$. From this and by virtue of Eq.\eqref{currents2} we 
readily calculate that $v_{\alpha \kappa}=-v_{\alpha -\kappa}$, which is in accordance with the arguments in \cite{Hanggi:2014}. 
For the overlap coefficients Eq.\eqref{currents2} yields $|C_{\alpha \kappa} (t)|^2=|C_{\alpha -\kappa} (T-t)|^2$, which ultimately 
gives 
\begin{equation}
 J(t_0)=-J(T-t_0) 
\end{equation}
from Eq.\eqref{currents1}. Analogously, the presence of parity symmetry induces $|C_{\alpha \kappa} (t)|^2=|C_{\alpha -\kappa} (t+T/2)|^2$. For the asymptotic currents this results in   
\begin{equation}
 J(t_0)=-J(t_0+T/2). 
\end{equation}
Note that both in the presence of time reversal- or parity symmetry the asymptotic current averaged over the initial time vanishes, i.e. $\tilde{J}\equiv1/T \int_0^T  J(t) \; dt=0$. 
  
\subsection{Transport in the spatiotemporally driven lattice}
In the following we demonstrate transport phenomena in the spatiotemporally driven lattice. As an exemplary setup we again consider a lattice with three barriers 
in a unit cell, i.e. $n_p=3$. The phases of the barriers are $(\delta_1, \delta_2, \delta_3)$ with $\delta_1=0$ and $\delta_2=2\pi/3$.\\
We calculate the asymptotic quantum current $J(t_0)$ numerically for different values of the third barrier phase $\delta_3$ and for different initial times $t_0$.
The results are shown in Fig. \ref{fig:transport} (a). For the case of $\delta_3=0$ the Hamiltonian possesses parity symmetry and thus according to the previous section the asymptotic current obeys
$J(t_0)=-J(t_0+T/2)$. If $\delta_3$ deviates from zero, parity symmetry is absent which allows for a nonzero averaged current. Interestingly, for the shown small values of $\delta_3$,  
it seems that the curve $J(t_0)$ is merely shifted by some constant while the overall shape is approximately independent of $\delta_3$. 
This also matches the observation concerning the variation of the Floquet spectrum for small deviations from a parity symmetric setup (cf. Fig. \ref{fig:spectrum} (a)). 
In this case, 
we observed that the spectrum approximately maintains its overall form, but
the symmetry axis of the spectrum, which is at $\kappa=0$ for $\delta_3=0$, is shifted to a nonzero $\kappa$ for small nonzero $\delta_3$.  In order to understand how these two observations could be related,
let us consider only the most populated Floquet mode for our initial state of a Gaussian wave packet, which for a parity symmetric setup with $\delta_3=0$ has zero quasi-momentum. 
As a crude approximation, one can argue that for a small deviation from $\delta_3=0$ this mode will remain almost unchanged but is shifted to a nonzero quasi-momentum in the same way that the spectrum is. 
Hence, the mode picks up a momentum which is related to the shift of the spectrum and thus related to the value of $\delta_3$. Furthermore, the spectrum is of course independent of the initial time and 
consequently the shift of the spectrum is the same for every $t_0$, which by virtue of the previous arguments would explain why the increase of $J(t_0)$ for increasing $\delta_3$ was to a good approximation independent of        
$t_0$.\\
For larger values of the third barrier phase $\delta_3$ the simple picture of a merely shifted spectrum is not longer applicable. 
The current averaged over the initial time is shown over the entire range of $\delta_3$ in Fig. \ref{fig:transport} (b) revealing the more complex behaviour at larger $\delta_3$, such as several sign changes 
of the currents which can not be explained easily be means of simple symmetry arguments. In fact it was shown in \cite{Denisov:2007} that the asymptotic quantum current as a function of some system parameter generally 
features a highly nontrivial dependence.

\section{Conclusion and Outlook}
\label{S6}

We have investigated the setup of a quantum particle in a periodic lattice consisting of driven Gaussian barriers. Since we allowed for different driving laws which were spatially periodically repeated, we have been able 
to design lattices with complex unit cells containing differently driven barriers. Within the framework of Floquet theory, we
presented an efficient numerical scheme that provided us with
the Floquet-Bloch modes for arbitrary quasi-momentum. 
For the Floquet spectrum we found that
the site-dependent driving has remarkable ramifications. A small deviation from a uniform, i.e. site-independent, driving was shown to cast exact crossings into avoided ones while 
the quasi-energy bands away from the crossings remained approximately unaltered. The width of the respective avoided crossings could be manipulated 
by adjusting parameters of the driving law.        
Because the presence of exact and avoided crossings
in the Floquet spectrum of driven lattice systems has been shown in the literature to be at the heart of many interesting phenomena, such 
as resonances in directed particle motion \cite{Denisov:2007}, the diffusion of a wave packet \cite{Graham:1994},  the stimulation of Landau-Zener transitions 
via external forces \cite{Flach:2007} or even the possibility of an absolute negative mobility \cite{Weitz:2013},
this control over the crossing's widths as achieved in this work opens up a promising direction of interesting future research. 
We could explain the effect of a crossing- to avoided crossing transition as the result of a breaking of the translational symmetry over the distance of two adjacent barriers. Since this effect was shown to be
symmetry induced it does in no way 
depend on the fine tuning of parameters and should be accessible in state of the art cold atom experiments. 
Promising experimental techniques for the realization of the required breaking of translational invariance are sub-wavelength lattices where modulations below the laser's
wavelength can be obtained \cite{Weitz:2009}. Possible setups which should allow for the phase modulated driving as studied in this work are provided by so called 'painted potentials' where the full 
control over the motion of each potential barrier is achieved \cite{Boshier:2009}.
Here we showed that directed particle motion can be generated even in situations where each barrier on its own 
does not break the necessary symmetries. Over a certain regime the resulting currents were shown to be controllable by engineering the asymmetry of the Floquet spectrum through parameter variations 
of the site-dependent driving.

\section*{Acknowledgments}

We thank H. Schanz for helpful discussions.

\appendix
\section{Powers of the Floquet operator}
\label{A1}
The quantities of interest are the matrix elements of the powers of $H_f(x,t)$, i.e. we need to compute  $\langle \bra{\mu_{\kappa} n} H_f^p \ket{\nu_{\kappa} m} \rangle$. For $p=0$ 
one simply gets $\langle \bra{\mu_{\kappa} n} \mathbf{1} \ket{\nu_{\kappa} m} \rangle = \delta_{\mu \nu}\delta_{n m}$ due to the orthonormality of the basis vectors. For $p=1$
we obtain 
\begin{equation}
\begin{aligned}
  \langle &\bra{\mu_{\kappa} n} H_f(x,t) \ket{\nu_{\kappa} m} \rangle   \\
   &=  \frac{1}{T} \int_0^T  dt   e^{-i n\omega t} \bra{\mu_{\kappa}} \left(H(x,t)-i \frac{\partial}{\partial t}\right) \ket{\nu_{\kappa}} e^{i m\omega t}\\
   &= \bra{\mu_{\kappa}} H^{(n-m)}(x)\ket{\nu_{\kappa}} + \bra{\mu_{\kappa}} m \omega \ket{\nu_{\kappa}} \delta_{\mu \nu}\\
   &= H_{\mu \nu}^{\kappa\, (n-m)} + m \omega \delta_{m n} \delta_{\mu \nu}.
\end{aligned}
\label{app1}
\end{equation}
Matrix representations of higher powers of $H_f(x,t)$ can be calculated recursively:
\begin{equation}
\begin{aligned}
  \langle &\bra{\mu_{\kappa} n} H^p_f(x,t) \ket{\nu_{\kappa} m} \rangle\\ 
  &= \langle \bra{\mu_{\kappa} n} H_f(x,t)  H^{p-1}_f(x,t) \ket{\nu_{\kappa} m} \rangle \\
  &=\sum_{\mu' n' } \langle \bra{\mu_{\kappa} n} H_f(x,t) \ket{\mu'_{\kappa} n'} \rangle \langle \bra{\mu'_{\kappa} n'} H^{p-1}_f(x,t) \ket{\nu_{\kappa} m} \rangle \\
  &=\sum_{\mu' n' }  (H_{\mu \mu'}^{\kappa\, (n-n')} + n \omega \delta_{n n'} \delta_{\mu \mu'})  \langle \bra{\mu'_{\kappa} n'} H^{p-1}_f(x,t) \ket{\nu_{\kappa} m} \rangle
\end{aligned}
\end{equation}
where we have used the completeness of the product basis as well as Eq.\eqref{app1}.    

\section{Fourier expansion of the potential}
\label{A2}
The Fourier coefficients of the potential $V^{(n)}(x)$ are to be represented in the basis of $\mathcal{R}$. Again we exploit that we can omit the $\kappa$-dependence 
for the operator of the potential energy and the calculation becomes:
\begin{equation}
\begin{aligned}
  V_{\mu \nu}^{(n)} 
		    &= \bra{\mu} V^{(n)}(x) \ket{\nu}\\
		    &= \frac{1}{L}\int_{-\infty}^{+\infty} dx V^{(n)}(x) e^{i\frac{2\pi}{L n_P}(\nu-\mu)x } \\
		    &= \frac{1}{L}\int_{-\infty}^{+\infty} dx \sum_{i=1}^{n_p}  V_{\text{SB}}^{(n)}(x-x_{0,i}) e^{in (\omega t_0 +\delta_i)} e^{i\frac{2\pi}{L n_P}(\nu-\mu)x }
\end{aligned}
\end{equation}
where we have used the expression for the Fourier coefficient from Eq.\eqref{fourier coeff} and as before $V_{\text{SB}}(x,t)$ is the potential of a single oscillating barrier. 
At this point we make use of the fact that we have restricted ourselves to setups in which the different Gaussian barriers have 
no significant overlap with one another. Hence we can exchange summation and integration and apply the coordinate transformation $\tilde{x}=x-x_{0,i}$ for each of the $n_p$ integrals:
\begin{equation}
\begin{aligned}
  V_{\mu \nu}^{(n)} &= \frac{1}{L} \sum_{i=1}^{n_p} \int_{-\infty}^{+\infty} dx V_{\text{SB}}^{(n)}(x-x_{0,i}) e^{in (\omega t_0 +\delta_i)} e^{i\frac{2\pi}{L n_P}(\nu-\mu)x } \\
		    &= \frac{1}{L} \sum_{i=1}^{n_p} \int_{-\infty}^{+\infty} d\tilde{x} V_{\text{SB}}^{(n)}(\tilde{x}) e^{in (\omega t_0 +\delta_i)} e^{i\frac{2\pi}{L n_P}(\nu-\mu)(\tilde{x}+ x_{0,i} )  } \\
		    &= \sum_{i=1}^{n_p} V_{\text{SB},\, \mu \nu}^{(n)}  e^{i(n (\omega t_0 +\delta_i)+ \frac{2\pi}{Ln_p}(\nu -\mu)x_{0,i})}.
\end{aligned}
\end{equation}

\section{Symmetry of the time evolution operator}
\label{A3}

First, we show that the symmetry given in Eq.\eqref{new symmetry} is equivalent to the relation: 
\begin{equation}
 U_{\mu \nu}^{\kappa,j}=(U_{\mu \nu}^{-\kappa,-j})^{\maltese}
\label{to show}
\end{equation}
for the matrix elements of the  time evolution operator over the $j$th time step as given by Eq.\eqref{time evo operator 2} (we omit the dependence on the initial time $t_0$ for the sake of clarity)
and we have adopted the notation $\left(U_{\mu \nu}^{\kappa,j}\right)^{\maltese} = U_{-\nu -\mu}^{\kappa,j} $ from \cite{Denisov:2007}. 
As mentioned in the main text, the time evolution operator over an entire period of the driving is obtained as the product of the $N$ 
operators $U_{\mu \nu}^{\kappa,j}$. 
For reasons of clarity, we will omit the momentum indices $(\mu,\nu)$ for the time being.
In analogy, the time evolution operator for the time interval $(t_1,t_2)$ is given by: 
\begin{equation}
 U^{\kappa}(t_2,t_1) =\prod_{j=j_1}^{j_2} U^{\kappa,j}
\end{equation}
where $j_1$ and $j_2$ are the number of time steps corresponding to $t_1$ and $t_2$ and are given by $j_{1,2}=\lceil N/T \cdot t_{1,2}\rceil$.
Now, if we assume that Eq.\eqref{to show} holds we find:
\begin{equation}
\begin{aligned}
 U^{\kappa}(T-t_1,T-t_2) &= \prod_{j=N-j_2}^{N-j_1} U^{\kappa,j} = \prod_{j=j_2}^{j_1} U^{\kappa,N-j} \\
 &= \prod_{j=j_2}^{j_1} (U^{-\kappa,j})^{\maltese} =\left(\prod_{j=j_1}^{j_2} U^{-\kappa,j} \right)^{\maltese}\\
 &= (U^{-\kappa}(t_2,t_1))^{\maltese}.
\end{aligned}
\end{equation}
Thus, the symmetry in Eq.\eqref{new symmetry} follows indeed from Eq.\eqref{to show}. However, the validity of Eq.\eqref{to show} remains to be shown.\\
To make some progress we restrict ourselves to the first order expansion of $U_{\mu \nu}^{\kappa,j}$ given by Eq. \eqref{time evo operator 2}:  
\begin{equation}
\begin{aligned}
\\
 U_{\mu \nu}^{\kappa,j} &= 
\sum_{n=-\infty}^{+\infty} e^{in\omega j \Delta t} \left( \delta_{\mu \nu}\delta_{n 0} + i \Delta t H^{\kappa,(n)}_{\mu \nu}\right)\\
&= \delta_{\mu \nu} + i \Delta t \sum_{n=-\infty}^{+\infty} e^{in\omega j \Delta t} H^{\kappa,(n)}_{\mu \nu}
\end{aligned}
\end{equation}
where we have used Eq.\eqref{app1} for the matrix elements of the Floquet operator $H_f(x,t)$.
In comparison we obtain for $(U_{\mu \nu}^{-\kappa,-j})^{\maltese}$:
\begin{equation}
\begin{aligned}
(U_{\mu \nu}^{-\kappa,-j})^{\maltese} 
&= U_{-\nu -\mu}^{-\kappa,-j}\\ 
&= \delta_{-\nu -\mu} + i \Delta t \sum_{n=-\infty}^{+\infty} e^{in\omega (-j) \Delta t} H^{\kappa,(n)}_{-\nu -\mu}\\
&= \delta_{\mu \nu} + i \Delta t \sum_{n=-\infty}^{+\infty} e^{in\omega j \Delta t} H^{\kappa,(-n)}_{-\nu -\mu}\\
&=\delta_{\mu \nu} + i \Delta t \sum_{n=-\infty}^{+\infty} e^{in\omega j \Delta t} H^{\kappa,(n)}_{\mu \nu} \\
&= U_{\mu \nu}^{\kappa,j}
\end{aligned}
\end{equation}
Here we exploited that $H^{\kappa,(-n)}_{-\nu -\mu}=H^{\kappa,(n)}_{\mu \nu}$, which follows directly from Eq.\eqref{fourier hamiltonian}, 
as well as from the fact that the Fourier components of a time reversal symmetric function obey $V^{(n)}(x)=V^{(-n)}(x)$.


\end{document}